\documentclass[psfig,epsf]{article}
\usepackage{epsfig} \usepackage{amssymb} \usepackage{amsfonts}

\def\beq{\begin{equation}}
\def\eeq{\end{equation}}
\def\bea{\begin{eqnarray}}
\def\eea{\end{eqnarray}}
\def\bem{\begin{math}}
\def\eem{\end{math}}
\def\bit{\begin{itemize}}
\def\eit{\end{itemize}}
\def\bla{\begin{flushright}}
\def\ela{\end{flushright}}
\def\qq2{$Q^2$}               
\def\aa1{$A_1(x,Q^2)$}        
\def\ff1{$F_1(x,Q^2)$}        
\def\gg1{$g_1(x,Q^2)$}        
\newcommand{\z}{&\hspace*{-8pt}}

\begin{document}

\begin{center}
{\Large \bf The Differential Equation Method:\\
evaluation of 
complicated 
Feynman diagrams
} \\

\vspace{4mm}

A.V.Kotikov\\
Particle Physics Laboratory,
Joint Institute for Nuclear Research \\
141980 Dubna, Russia.\\
\end{center}

\begin{abstract}

We discuss a progress in calculation of Feynman integrals which has been
done with help of the Differential Equation Method and 
demonstrate the results for a class of two-point two-loop diagrams.
\end{abstract}

\vskip 1.5cm
The idea of the Differential Equation Method 
(see \cite{DEM1}-\cite{DEM3}) (see a reviews in \cite{DEMrev}): 
to apply the integration by parts procedure \cite{IntPart} to an internal 
$n$-point subgraph
of a complicated Feynman diagram and later to represent new complicated
diagrams, obtained here, as derivatives in respect of corresponding masses 
of the initial diagram.

The integration by parts procedure \cite{IntPart},
\cite{DEM1}-\cite{DEM3}
 for a general $n$-point (sub)graph with masses of its lines
$m_1, m_2, ..., m_n$, line momenta $p_1, p_2=p_1 - p_{12}, p_n=p_1 - p_{1n}$
and indices $j_1, j_2, .., j_n$, respectively, has the following form:
\begin{eqnarray}
0&=& \int d^Dp_1 \frac{\partial}{\partial p_1^{\mu}}~ 
\Biggl\{p_1^{\mu} ~
{\biggl(\prod_{i=1}^{n}
c_i^{j_i} \biggr)}^{-1} \Biggr\} \label{1} \\ 
&=& \int d^Dp_1  {\biggl(\prod_{i=1}^{n}
c_i^{j_i} \biggr)}^{-1} \Biggl[ D - 2j_1 \biggl( 1- \frac{m_1^2}{c_1} \biggr) 
- \sum_{i=2}^{n} j_i 
 \biggl( 1-
\frac{m_1^2 + m_i^2 + p^2_{1i}-c_1}{c_i} \biggr)
\Biggr],
\nonumber
\end{eqnarray}
where
$c_k=p^2_k+m^2_k$ are the propagators of $n$-point (sub)graph. 

Because the diagram with the index $(j_i+1)$ of the propagator $c_i$ may be 
represented as the derivative (on the mass $m_i$), Eq.(\ref{1}) leads to 
the differential equations (in principle, to partial differential equations)
for the initial diagram (having the index $j_i$, respectively). This approach
which is based on
 the Eq.(\ref{1}) and allows to construct the (differential) relations
between diagrams has been named as Differential Equations Method (DEM).  
 For most
interested cases (where the number of the masses is limited)
these partial differential equations may be represented through
original differential equation\footnote{The example of the direct application
of the partial differential equation may be found in \cite{FJ1}.},
which is usually simpler to analyze.

Thus, we have got the differential equations for the initial diagram. The
inhomogeneous term contains only more simpler diagrams.
These simpler diagrams have more trivial topological structure 
and/or less number of loops \cite{DEM1} and/or ends \cite{DEM2,DEM3}.

Applying the procedure several times, we will
able to represent complicated Feynman integrals (FI) and their derivatives
(in respect of internal masses) through a set of quite simple well-known 
diagrams. 
Then, the results for the
complicated FI can be obtained by integration several times
of the known results for corresponding simple diagrams
\footnote{In calculations of real processes 
(essentially in the framework of Standard Model)
it is useful to use
the relation (1) (at least, at first steps of calculations)
to decrease the number of contributed diagrams (see
\cite{DEM1}-\cite{DEM3} and \cite{FTT} and references therein).}.

Sometimes it is useful (see \cite{Remiddi}) to use external momenta 
(or some their
functions) but not masses as parameters of integration.\\

\vskip 0.5cm

{\bf The recent progress in calculation of Feynman integrals with help 
of the DEM.} \\

{\bf 1.} The articles \cite{FKV1} and \cite{FKV}: \\

{\bf a)} The set of two-point two-loop FI with one- and two-mass
thresholds has been evaluated by DEM (see Fig.1).
The results are given on pages 2 and 3 and of some of them have been
known before (see 
\cite{FKV1}). The check of the results 
has been
done by Veretin programs (see discussions in 
\cite{FKV1,FKV2} and references therein).\\
   \begin{figure}[tb]
\unitlength=1mm
\begin{picture}(140,80)
  \put(0,80){%
   \epsfig{file=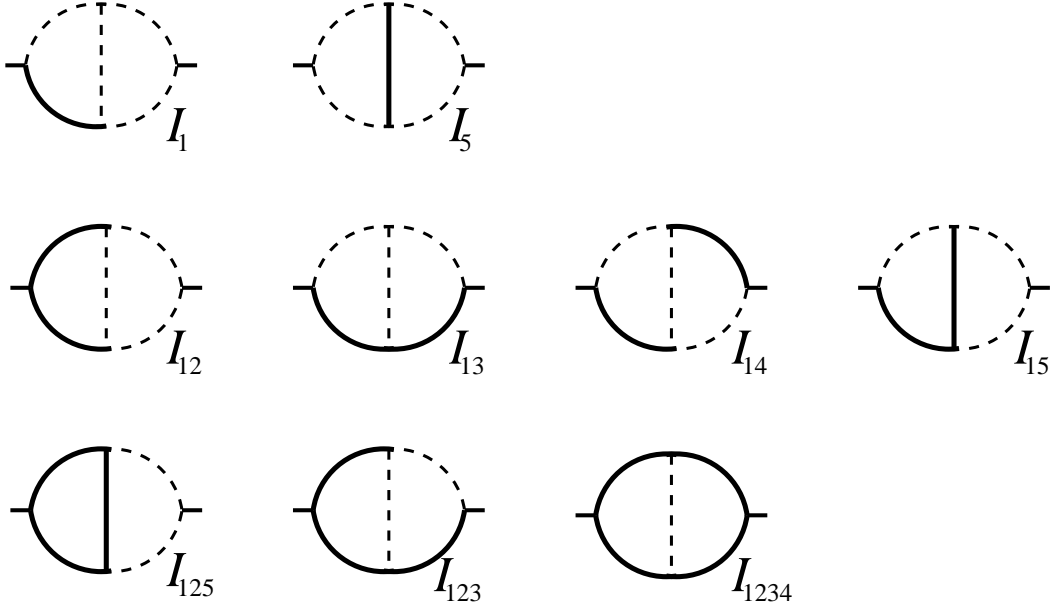,width=80mm,height=140mm,angle=-90}%
}
\end{picture}
 \caption{
Two-loop selfenergy diagrams.
Solid lines denote propagators with the mass $m$; dashed lines
denote massless propagators.}
\vskip 0.5cm
 \end{figure}

{\bf b)} The set of three-point two-loop FI with one- and two-mass
thresholds has been evaluated (the results of some of them has been
known before (see \cite{FKV1})) by a combination of DEM and Veretin programs
for calculation of first terms of FI small-moment expansion (see 
discussions in \cite{FKV1,FKV2} and references therein).\\

{\bf 2.} The article \cite{FKK}: \\

The full set of two-point two-loop onshell master diagrams 
has been evaluated by DEM. The check of the results has been
done by Kalmykov programs (see page 5 and discussions in 
\cite{FKK,FKK1} and references therein). \\

{\bf 3.} The articles \cite{GeRe}: \\

The set of three-point and four-point two-loop massless FI 
has been evaluated.\\

\vskip 0.5cm
{\bf Here we demonstrate the results of FI are displayed on Fig.1.}\\

We introduce the notation for
polylogarithmic functions \cite{Lewin}: 
\begin{eqnarray}
  {\rm Li}_a(z) = S_{a-1,1}(z), ~~~ 
S_{a+1,b}(z) = \frac{(-1)^{a+b}}{a!\,b!}
   \int_0^1 \frac{\log^a(t)\log^b(1-zt)}{t}\,dt.
\nonumber
\end{eqnarray}

  We introduce also the following two variables 
\begin{eqnarray}
z=\frac{q^2}{m^2},~~~y=\frac{1-\sqrt{z/(z-4)}}{1+\sqrt{z/(z-4)}}\,.
\nonumber
\end{eqnarray}

  Then\footnote{We would like to note that the
coefficients of expansions of the results (\ref{eq}) in respect of $z$ 
are very similar (see \cite{KaKo}) to results 
for the moments of structure functions of deep inelastic scattering.}

\begin{eqnarray}
q^2 \cdot I_1 \z=\z 
     - \frac{1}{2}\log^2(-z)\log(1-z) 
     - 2\log(-z){\rm Li}_2(z)
     + 3{\rm Li}_3(z) -6 S_{1,2}(z)  
- \log(1-z) \biggl(  \zeta_2 + 2{\rm Li}_2(z) \biggr)\,, \nonumber \\
\z\z\nonumber\\
q^2 \cdot I_5 \z=\z 
      2 \zeta_2 \log(1+z) + 2\log(-z){\rm Li}_2(-z) + \log^2(-z)\log(1+z)
        + 4\log(1+z){\rm Li}_2(z) 
            \nonumber \\
      \z-\z 2{\rm Li}_3(-z) 
           - 2{\rm Li}_3(z) 
           + 2 S_{1,2}(z^2) - 4S_{1,2}(z) - 4 S_{1,2}(-z)\,, \nonumber \\
\z\z\nonumber\\
q^2 \cdot I_{12} \z=\z
     {\rm Li}_3(z) - 6 \zeta_3 - \zeta_2\log y
     - \frac16\log^3 y - 4\log y\,{\rm Li}_2(y)  
+ 4{\rm Li}_3(y) - 3{\rm Li}_3(-y) + \frac13 {\rm Li}_3(-y^3)\,,\nonumber \\
\z\z\nonumber\\
q^2 \cdot I_{13} \z=\z 
      - 6 S_{1,2}(z)
      - 2 \log(1-z) \biggl(  \zeta_2 + {\rm Li}_2(z) \biggr)\,, \nonumber \\
\z\z\nonumber\\
q^2 \cdot I_{14} \z=\z  
     \log(2-z) \biggl( \log^2(1-z) -2 \log(-z)\log(1-z) -2{\rm Li}_2(z)
       \biggr)
-\frac{2}{3} \log^3(1-z) - 2 \zeta_2 \log(1-z)
\nonumber \\  \z+\z \log(-z)\log^2(1-z)
-S_{1,2}\bigr(1/(1-z)^2\bigl)
     + 2 S_{1,2}\bigr(1/(1-z)\bigl) + 2 S_{1,2}\bigr(-1/(1-z)\bigl)  
+ \frac13 \log^3 y
\nonumber\\
  \z+\z 
\log^2 y\,\biggl( 2 \log(1+y^2) -3 \log(1-y +y^2) \biggr) 
-6 \zeta_3 - {\rm Li}_3(-y^2)
     +\frac23 {\rm Li}_3(-y^3) - 6 {\rm Li}_3(-y) 
         \nonumber \\ 
\z+\z 2\log y\, \biggl( {\rm Li}_2(-y^2) -{\rm Li}_2(-y^3) 
              +3 {\rm Li}_2(-y)    \biggr)\,,  \nonumber \\
\z\z\nonumber\\
q^2 \cdot I_{15} \z=\z 
   2{\rm Li}_3(z) - \log(-z)\,{\rm Li}_2(z) 
              + \zeta_2 \log(1-z) 
   - \frac12\log^2 y\,\biggl( 8 \log(1-y) -3 \log(1-y +y^2) \biggr) 
-6 \zeta_3                    \nonumber \\
\z+\z  \frac16 \log^3 y 
- \frac13 {\rm Li}_3(-y^3) + 3 {\rm Li}_3(-y) 
            + 8 {\rm Li}_3(y) 
+\log y\, \biggl( {\rm Li}_2(-y^3) -3 {\rm Li}_2(-y) 
              -8 {\rm Li}_2(y)   \biggr)\,, \nonumber \\
\z\z\nonumber\\
q^2 \cdot I_{123} \z=\z 
    - \zeta_2 \biggl(  \log(1-z) +  \log y \biggr)
    - 6 \zeta_3  -\frac32 \log(1-y +y^2)\, \log^2 y 
+ {\rm Li}_3(-y^3) - 9 {\rm Li}_3(-y)
\nonumber \\
    \z-\z  2 \log y\, \biggl( {\rm Li}_2(-y^3) - 3{\rm Li}_2(-y) \biggr)  
      \,, \nonumber \\
\z\z\nonumber\\
q^2 \cdot I_{125} \z=\z  
      -2 \log^2 y\, \log(1-y)
      - 6 \zeta_3 + 6 {\rm Li}_3(y) - 6\log y\, {\rm Li}_2(y)\,, \nonumber \\
\z\z\nonumber\\
q^2 \cdot I_{1234} \z=\z 
    - 6 \zeta_3  - 12 {\rm Li}_3(y) - 24 {\rm Li}_3(-y)
+8 \log y\, \biggl( {\rm Li}_2(y) + 2{\rm Li}_2(-y) \biggr)  
      + 2 \log^2 y\, \biggl( \log(1-y) 
             \nonumber \\    \z+\z 
 2 \log(1+y) \biggr)\,. 
\label{eq}
\end{eqnarray}


\vskip 0.5cm
{\bf Here we demonstrate the results of FI are displayed on Fig.2.}\\

\begin{figure}[tb]
\vskip -0.5cm
\epsfig{figure=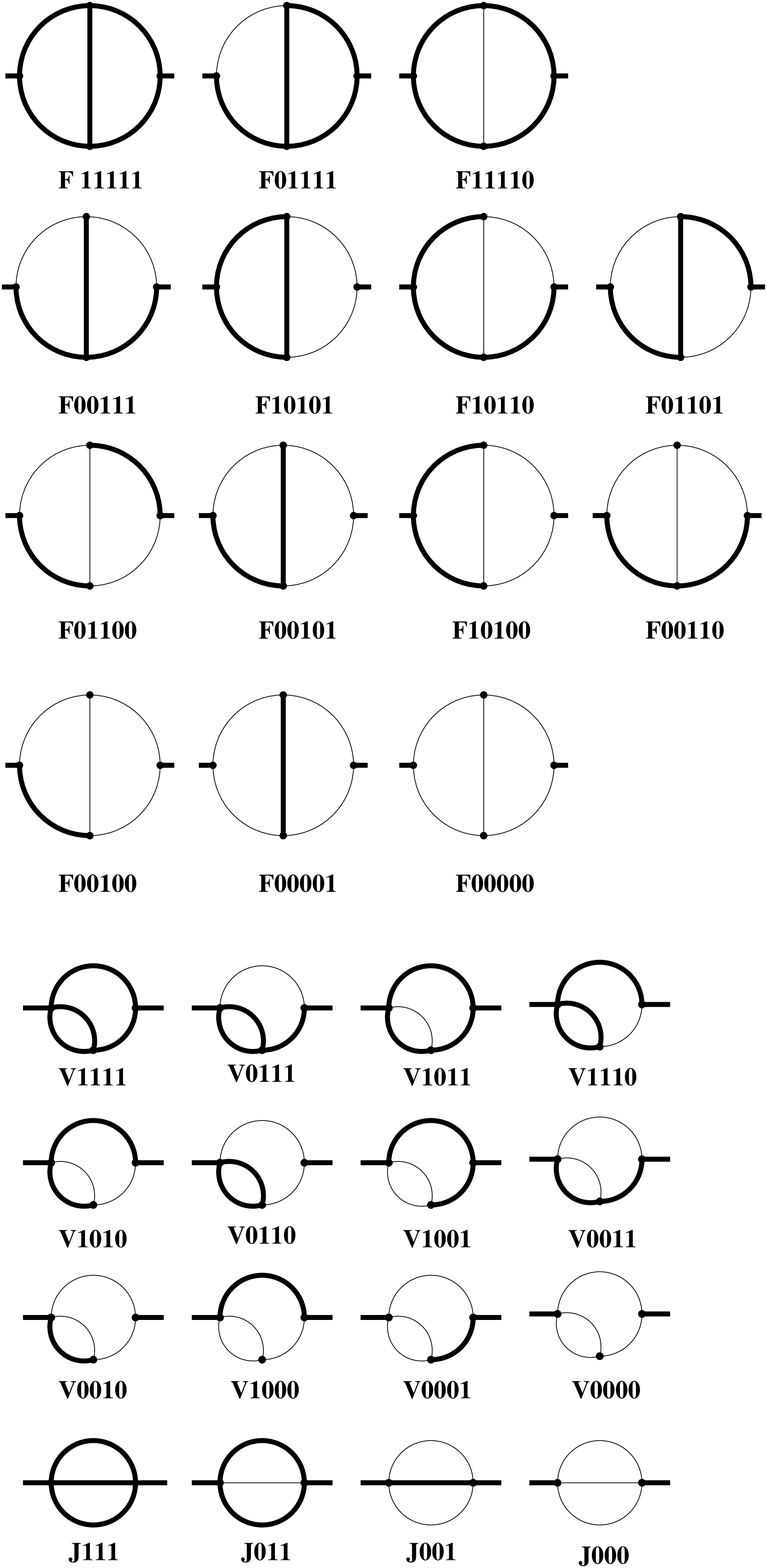,height=7.3in,width=5.2in}
\caption{\label{set} The full set of two-loop self-energies diagrams
with one mass.  Bold and thin lines correspond to the mass and
massless propagators, respectively.}
\end{figure}

We consider here the following master-integrals in Euclidean 
space-time with dimension $D= 4-2\varepsilon$:

\begin{eqnarray}
 {\bf ONS} \{ {\cal IJ} \} (i,j,m)   & \equiv & K^{-1} 
\int d^Dk P^{(i)}(k,{\cal I}m) 
\left. P^{(j)}(k-p,{\cal J}m) \right|_{p^2=-m^2},
\nonumber \\
{\bf J} \{ {\cal IJK} \} (i,j,k,m)  & \equiv & K^{-2}
\int  d^Dk_1 d^Dk_2
P^{(i)}(k_1,{\cal I} m) P^{(j)}(k_1-k_2,{\cal J}m)
\left. P^{(k)}(k_2-p,{\cal K}m) \right|_{p^2=-m^2},
\nonumber  \\
{\bf V} \{ {\cal IJKL} \}  (i,j,k,l,m)  & \equiv & K^{-2}
\int  d^Dk_1 d^Dk_2 P^{(i)}(k_2-p,{\cal I}m) 
\nonumber \\
&& 
\times 
P^{(j)}(k_1-k_2,{\cal J}m) P^{(k)}(k_1,{\cal K}m)
\left. P^{(l)}(k_2,{\cal L}m) \right|_{p^2=-m^2},
\nonumber  \\
{\bf F}\{ {\cal ABIJK} \}  (a,b,i,j,k,m)  
& \equiv & m^2 K^{-2}
\int  d^Dk_1 d^Dk_2
P^{(a)}(k_1,{\cal A} m) P^{(b)}(k_2,{\cal B} m)
\nonumber \\
&&
\times
P^{(i)}(k_1-p,{\cal I}m)
P^{(j)}(k_2-p,{\cal J}m)
\left. P^{(k)}(k_1-k_2,{\cal K}m) \right|_{p^2=-m^2},
\nonumber 
\end{eqnarray}

\noindent
where $$
K = \frac{\Gamma(1+\varepsilon)}{ \left(4 \pi \right)^{\frac{D}{2}}
\left( m^2 \right)^{\varepsilon}}, ~~~ 
P^{(l)}(k,m) \equiv \frac{1}{(k^2+m^2)^l},$$ 
the normalization factor $1/(2 \pi)^D$ for each loop is assumed,
and ${\cal A,B,I,J,K} = 0,1.$

The finite part of most of the F-type master-integrals can be obtained 
from results of Ref.\cite{FKV1} in the limit $z \rightarrow 1$. 
{\bf F10101} and 
{\bf F11111} have been calculated in Refs.\cite{2david,f11111}, respectively.
Instead of the usually taken {\bf F01101} integral \cite{2david,1david} 
we consider  {\bf J111} as master integral. 
We recall the results of all master integrals for completeness.
The last 
master integral 
{\bf F00111} has been found in \cite{FKK}.

The finite part of the integrals of V- and J-type can be found in 
Refs.\cite{short}.
The calculation of the $\varepsilon$ e
($\varepsilon^2$) parts of master integrals of this type have been 
performed by DEM.

The results for F-type master-integrals are follows:

\begin{equation}
{\bf F}\{ {\cal ABIJK} \} (1,1,1,1,1,m) = a_1 \zeta(3) 
+ a_2 \frac{\pi}{\sqrt{3}} S_2 + a_3 i \pi \zeta(2) + {\cal O} (\varepsilon),
\label{first}
\end{equation}

\noindent
and the coefficients $\{a_i \}$ are given in Table I:

$$
\begin{array}{|c|c|c|c|c|c|c|c|c|} \hline
\multicolumn{9}{|c|}{TABLE ~~~I}    \\   \hline
& {\bf F11111} & {\bf F00111} & {\bf F10101} & {\bf F10110} &
  {\bf F01100} & {\bf F00101} & {\bf F10100} & {\bf F00001}
\\[0.3cm] \hline
a_1 & -1 & 0 & -4 & -1 & 0 & -3 & -2 & -3 \\[0.3cm] \hline
a_2 & \frac{9}{2} & 9 & \frac{27}{2} & 9 & \frac{27}{2} &
\frac{27}{2} & 9 & 0 \\[0.3cm] \hline
a_3 & 0 & 0 & \frac{1}{3} & 0 & 1 & 1 & \frac{2}{3} & 1
\\[0.3cm] \hline
\end{array}
$$

\noindent
where \cite{S-M,2david,Lewin} 
$$S_2 = \frac{4}{9\sqrt{3}} {\rm Cl}_2 \left(\frac{\pi}{3} 
\right)=0.260434137632 \cdots.$$ 
Here we used the $m^2-i \varepsilon$ prescription. The results for
the remaining master integrals are the following ones:
\begin{eqnarray}
{\bf V}\{ {\cal IJKL} \} (1,1,1,1,m) &=& 
\frac{1}{2 \varepsilon^2}
+ \frac{1}{\varepsilon}
\left( \frac{5}{2} - \frac{\pi}{\sqrt{3}} \right)
+ \frac{19}{2} + \frac{b_1}{2} \zeta(2) -4 \frac{\pi}{\sqrt{3}}
- \frac{63}{4} S_2 + \frac{\pi}{\sqrt{3}} \ln 3 
\nonumber \\ 
&+& \hskip -8pt
 \varepsilon \Biggl\{ 
\frac{65}{2} + b_2 \zeta(2) - b_3 \zeta(3)
- 12 \frac{\pi}{\sqrt{3}} - 63 S_2 + b_4 \zeta(2) \ln3
+ \frac{9}{4} b_4 S_2 \frac{\pi}{\sqrt{3}}
\nonumber \\ 
&+& \hspace*{-8pt}
\frac{63}{4} S_2 \ln3
+ 4 \frac{\pi}{\sqrt{3}} \ln 3
- \frac{1}{2} \frac{\pi}{\sqrt{3}} \ln^2 3
- \frac{b_5}{2} \frac{\pi}{\sqrt{3}} \zeta (2)
- \frac{21}{2} \frac{{\rm Ls}_3 \left(\frac{2\pi}{3} \right)}{\sqrt{3}}
\Biggr\}
+ {\cal O} (\varepsilon^2),
\nonumber
\end{eqnarray}

\noindent
where the coefficients $\{b_i \}$ are listen in Table II
\footnote{The results for the master integral {\bf V1001} had a little
error (see \cite{DaKa})}:

$$
\begin{array}{|c|c|c|c|c|c|} \hline
\multicolumn{6}{|c|}{TABLE ~~~II}    \\   \hline
            &  b_1 & b_2  & b_3 & b_4 & b_5  \\[0.3cm] \hline
{\bf V1111} &  - 1 & -6 & \frac{9}{2}   & 4 & 9   \\[0.3cm] \hline
{\bf V1001} &    3 &  8 & -\frac{3}{2} & 0 & 21  \\[0.3cm] \hline
\end{array}
$$

\begin{eqnarray}
{\bf J111}(1,1,1,m) &=& - m^2 \Biggl(
\frac{3}{2\varepsilon^2} + \frac{17}{4 \varepsilon} + \frac{59}{8}
+ \varepsilon \Biggl\{\frac{65}{16} + 8 \zeta(2) \Biggr \}
\nonumber \\
&-& 
-\varepsilon^2 \Biggl\{
\frac{1117}{32} - 52 \zeta(2) + 48 \zeta (2) \ln 2 - 28 \zeta (3)
\Biggr\} + {\cal O} (\varepsilon^3)
\Biggr), 
\label{master-j111} \\
%
& & \nonumber \\
{\bf J011}(1,1,2,m) &=& \frac{1-4\varepsilon}{2 (1-2\varepsilon) (1-3\varepsilon)}
\Biggl( \frac{1}{\varepsilon^2} + 2 \frac{\pi}{\sqrt{3}}
- \frac{2}{3} \zeta(2) 
\nonumber \\
&+& \varepsilon \Biggl\{
8 \frac{\pi}{\sqrt{3}} - \frac{2}{3} \zeta(2)
-6  \frac{\pi}{\sqrt{3}} \ln 3
+ \frac{2}{3} \zeta(3)
+ 27 S_2 \Biggr\}  
+ {\cal O} (\varepsilon^2)
\Biggr),
\label{master-j011-1}\\
& & \nonumber \\
{\bf J011}(1,1,1,m) &=& -  \frac{m^2}{2} 
\frac{4- 15 \varepsilon}{(1-2\varepsilon) (1-3\varepsilon) (2-3\varepsilon)}
\Biggl( \frac{1}{\varepsilon^2} + \frac{3}{2} \frac{\pi}{\sqrt{3}}
+ \varepsilon \Biggl\{
\frac{45}{8} \frac{\pi}{\sqrt{3}} 
-  \frac{9}{2}  \frac{\pi}{\sqrt{3}} \ln 3
+ \frac{81}{4} S_2 \Biggr\}  \nonumber \\
&+&
 \varepsilon^2 \Biggl\{
12 - \zeta(2) - \frac{1863}{16} S_2 -\frac{867}{32} \frac{\pi}{\sqrt{3}} 
+ \frac{207}{8} \frac{\pi}{\sqrt{3}} \ln 3 
+ \frac{243}{4}S_2 \ln 3
\nonumber \\
&-& 
\frac{27}{4} \frac{\pi}{\sqrt{3}} \ln^2 3
- 21 \frac{\pi}{\sqrt{3}} \zeta(2)
- \frac{81}{2}  \frac{{\rm Ls}_3 \left(\frac{2\pi}{3} \right)}{\sqrt{3}} 
\Biggr\} +
{\cal O} (\varepsilon^3)
\Biggr),
\label{master-j011-2}\\
& & \nonumber \\
{\bf ONS11}(1,1,m) &=&
\frac{1}{1-2\varepsilon} \Biggl[
\frac{1}{\varepsilon}  - \frac{\pi}{\sqrt{3}}
+ \varepsilon \left\{ \frac{\pi}{\sqrt{3}}\ln3 - 9 S_2 \right\}
\nonumber \\
&+& \varepsilon^2 \left\{
9 S_2 \ln3 - \frac{1}{2}\frac{\pi}{\sqrt{3}} \ln^2 3
- 6 \frac{{\rm Ls}_3 \left(\frac{2\pi}{3} \right)}{\sqrt{3}}
- 3 \frac{\pi}{\sqrt{3}} \zeta (2) \right\}
+ {\cal O} (\varepsilon^3) \Biggr],
\label{master-ons11}
\end{eqnarray}

\noindent
where \cite{Lewin}  
$$ {\rm Ls}_3(x) = -\int_0^x \ln^2 
\left| 2 \sin \frac{\theta}{2}\right| d \theta ~~\mbox{ and }~~
{\rm Ls}_3 \left(\frac{2\pi}{3} \right) = -2.14476721256949 \cdots$$

   The above results were checked numerically. Pad\'e approximants 
were calculated from the small momentum Taylor expansion of the
diagrams \cite{small}. The Taylor coefficients were obtained by means of the
package \cite{TLAMM} with the master integrals taken from \cite{small1}.
Further we made use 
of the idea of Broadhurst \cite{numeric} to apply the FORTRAN program 
{\bf PSLQ}
\cite{pslq} to express the obtained numerical values in terms of
a `basis' of irrational numbers, which were predicted by  DEM.

Let us point out that the numbers we obtain are related to
polylogarithms at the sixth root of unity\footnote{For 
the results obtained in $1/N$ expansion, however, the 
arguments of polylogarithms 
have other values (see \cite{BGK}).}
and hence are in the same class of
transcendentals obtained by Broadhurst \cite{numeric}
in his investigation of three-loop diagrams
which correspond to a closure of the two-loop topologies considered here.\\

  {\it Acknowledgments.}
Author
 would like to express his sincerely thanks to the Organizing
 Committees of the Research Workshop ``Calculations for modern and future
Colliders'' and the XVth International Workshop ``High Energy Physics
and Quantum Field Theory''
and especially to E.E. Boss, V.A. Ilyin, D.I. Kazakov and D.V. Shirkov
 for the kind invitation, 
the financial support
at  such remarkable Conferences, and 
for fruitful discussions.
Author
was supported in part
by Alexander von Humboldt fellowship. 

%

%

\end{document}